\documentclass{appolb}
\usepackage{epsfig}
\newcommand{\beq}[1]{
\begin{equation}\label{#1}}
\newcommand{\eeq}{\end{equation}}
\newcommand{\bea}[1]{
\begin{eqnarray}\label{#1}}
\newcommand{\eea}{\end{eqnarray}}
\begin{document}
\pagestyle{plain}
\eqsec
\newcount\eLiNe\eLiNe=\inputlineno\advance\eLiNe by -1
\title{NLO BFKL at work: the electroproduction of two light vector mesons
\thanks{Presented by A. Papa at the ``School on QCD, Low-$x$ Physics, 
Saturation and Diffraction'', Copanello (Calabria, Italy), July 1 - 14, 2007.}}
\author{Dmitry Yu. Ivanov
\address{Sobolev Institute of Mathematics \\
630090 Novosibirsk, Russia}
\and
Alessandro Papa
\address{Dipartimento di Fisica, Universit\`a della Calabria \\
and Istituto Nazionale di Fisica Nucleare, Gruppo collegato di Cosenza \\
I-87036 Arcavacata di Rende, Cosenza, Italy}}
\maketitle

\begin{abstract}
The forward electroproduction of two light vector mesons is the first example of a
collision process between strongly interacting colorless particles for
which the amplitude can be written completely within perturbative QCD in the Regge
limit with next-to-leading accuracy. This amplitude can be written as a 
convolution of two impact factors for the virtual photon to light vector meson 
transition with the BFKL Green's function. 
In this lecture we first describe how the relevant impact factor is calculated,
then we perform the convolution with the BFKL Green's function and illustrate
the numerical procedure to obtain a well-behaved amplitude.
\end{abstract}

\section{Introduction}

\begin{figure}[tb]
\begin{center}
{\epsfysize 5cm \epsffile{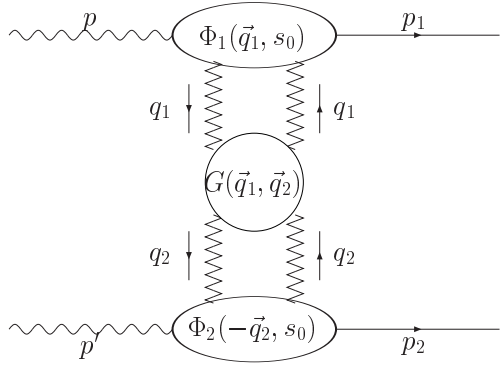}}
\caption[]{Schematic representation of the amplitude for the $\gamma^*(p)\,
\gamma^*(p') \to V(p_1)\, V(p_2)$ forward scattering.}
\end{center}
\label{fig:BFKL}
\end{figure}

In the BFKL approach~\cite{BFKL}, both in the leading logarithmic approximation 
(LLA), which means resummation of leading energy logarithms, all terms 
$(\alpha_s\ln(s))^n$, and in the next-to-leading approximation (NLA), which means 
resummation of all terms $\alpha_s(\alpha_s\ln(s))^n$, the (imaginary part of the) 
amplitude for a large-$s$ hard collision process can be written as the convolution 
of the Green's function of two interacting Reggeized gluons with the impact factors
of the colliding particles (see, for example, Fig.~\ref{fig:BFKL}).

The Green's function is determined through the BFKL equation. The NLA singlet
kernel of the BFKL equation has been achieved in the forward
case~\cite{NLA-kernel}, after the long program of calculation of the NLA
corrections~\cite{NLA-corrections} (for a review, see Ref.~\cite{news}).
For the non-forward case the ingredients to the NLA BFKL kernel are known since
a few years for the color octet representation in the
$t$-channel~\cite{NLA-corrections-nf}. This color representation is very
important for the check of consistency of the $s$-channel unitarity with
the gluon Reggeization, i.e. for the ``bootstrap''~\cite{FF98,bootstrap}.
Recently it was completed also the calculation of the non-forward NLA BFKL kernel 
in the singlet color representation, i.e. in the Pomeron channel, relevant 
for physical applications~\cite{FF05}.

On the other side, NLA impact factors have been calculated for colliding
partons~\cite{FFKP99,partonIF} and for forward jet production~\cite{BCV03}.
Among the impact factors for transitions between colorless objects, the most
important one from the phenomenological point of view is certainly the
impact factor for the virtual photon to virtual photon transition, i.e.
the $\gamma^* \to \gamma^*$ impact factor, since it would open the way to
predictions of the $\gamma^* \gamma^*$ total cross section. Its calculation is
rather complicated and it was completed after year-long 
efforts~\cite{gammaIF,FIK,FIK1}.

A considerable simplification can be gained if one considers instead the impact
factor for the transition from a virtual photon $\gamma^*$ to a light
neutral vector meson $V=\rho^0, \omega, \phi$. In this case, indeed, a close
analytical expression can be achieved in the NLA, up to contributions
suppressed as inverse powers of the photon virtuality~\cite{IKP04}. In particular,
it turns out that (a) the dominant helicity amplitude is that for the transition
from longitudinally polarized virtual photon to longitudinally polarized
vector meson; (b) the impact factor, both in the LLA and in the NLA, factorizes
into the convolution of a hard scattering amplitude, calculable in perturbative
QCD, and a meson twist-2 distribution amplitude~\cite{IKP04}.

The knowledge of the $\gamma^* \to V$ impact factor allows for the first time
to determine completely within perturbative QCD and with NLA accuracy the amplitude
of a physical process, the $\gamma^* \gamma^* \to V V$ 
reaction~\cite{IP06,IP07}. This possibility
is interesting first of all for theoretical reasons, since it can shed light
on the role and the optimal choice of the energy scales entering the BFKL approach.
Moreover, it can be used as a test-ground for comparisons with approaches
different from BFKL, such as DGLAP, and with possible next-to-leading order
extensions of phenomenological models, such as color dipole and $k_t$-factorization.
But it could be interesting also for the possible applications to the
phenomenology. Indeed, the calculation of the $\gamma^* \to V$ impact factor is
the first step towards the application of BFKL approach to the description of
processes such as the vector meson electroproduction $\gamma^* p\to V p$, being
carried out at the HERA collider, and the production of two mesons in the photon
collision, $\gamma^*\gamma^*\to VV$ or $\gamma^* \gamma \to VJ/\Psi$, which can be
studied at high-energy $e^+e^-$ and $e\gamma$ colliders.

In this paper we concentrate on the NLA forward amplitude for the
$\gamma^* \gamma^* \to V V$ reaction (Section~2). Such a process has been studied
in Ref.~\cite{PSW} in the Born (2-gluon exchange) limit for arbitrary transverse 
momentum and, for the forward case only, in Ref.~\cite{PSW1} with LLA plus an 
estimate of NLA effects.\footnote{The QCD factorization properties of this amplitude
have been studied In Ref.~\cite{BSSW06}.}

First of all, we show how the available results for the $\gamma^* \to V$ impact factor
(Section~3) and the BFKL Green's function can be put together to build up the 
NLA amplitude of the $\gamma^* \gamma^* \to V V$ process in the $\overline {\mbox{MS}}$ 
scheme (Section~4). Then we restrict ourselves to the particular case of
collision of virtual photons with equal virtualities and present some numerical
estimates of our result, aimed at showing the extent of the contributions to the
NLA amplitude from the impact factor and from the NLA kernel and the dependence on
the energy scale introduced in the BFKL approach and on the renormalization scale
which appears in the $\overline {\mbox{MS}}$ scheme. We show that, despite being the 
NLA corrections large and of opposite sign with respect to the leading order, it 
is possible to achieve a well-behaved form of the amplitude, by a suitable choice 
of the energy and renormalization scale parameters (Section~5).

Then, we compare different procedures to optimize the perturbative result and
different representations of the amplitude, in order to have an estimate of the 
systematic effects which underlie our determination (Section~6).
Finally, we calculate the differential cross section at the minimum squared 
momentum transfer and compare it with the approach of Ref.~\cite{PSW1} (Section~7).

The use in our approach of the BFKL kernel improved by the inclusion of subleading 
terms generated by renormalization group analysis, which has been suggested to cure 
the instabilities in the behavior of the BFKL Green's function in the next-to-leading 
approximation~\cite{Sal98}, has been studied in Ref.~\cite{CPS07} and is presented
in Ref.~\cite{Cap08}. The use of such an improvement has allowed for the numerical 
determination of the amplitude also in the case of colliding photons with strongly
ordered virtuality.

\section{The amplitude for the electroproduction of two light vector mesons:
kinematics and BFKL structure}

We consider the production of two light vector mesons ($V=\rho^0, \omega, \phi$) in
the collision of two virtual photons,
\beq{process}
\gamma^*(p) \: \gamma^*(p')\to V(p_1) \:V(p_2) \;.
\eeq
Here, $p_1$ and $p_2$ are taken as Sudakov vectors satisfying $p_1^2=p_2^2=0$ and
$2(p_1 p_2)=s$; the virtual photon momenta are instead
\beq{kinphoton}
p =\alpha p_1-\frac{Q_1^2}{\alpha s} p_2 \;, \;\;\;\;\;
p'=\alpha^\prime p_2-\frac{Q_2^2}{\alpha^\prime s} p_1 \;,
\eeq
so that the photon virtualities turn to be $p^2=-Q_1^2$ and $(p')^2=-Q_2^2$.
We consider the kinematics when
\beq{kin}
s\gg Q^2_{1,2}\gg \Lambda^2_{QCD} \, ,
\eeq
and
\beq{alphas}
\alpha=1+\frac{Q_2^2}{s}+{\cal O}(s^{-2})\, , \quad
\alpha^\prime =1+\frac{Q_1^2}{s}+{\cal O}(s^{-2})\, .
\eeq 
In this case vector mesons are produced by longitudinally polarized photons in
the longitudinally polarized state~\cite{IKP04}. Other helicity amplitudes are
power suppressed, with a suppression factor $\sim m_V/Q_{1,2}$.
We will discuss here the amplitude of the forward scattering, i.e.
when the transverse momenta of produced $V$ mesons are zero or 
when the variable $t=(p_1-p)^2$ takes its maximal value $t_0=-Q_1^2Q_2^2/s+{\cal
O}(s^{-2})$.

The forward amplitude in the BFKL approach may be presented as follows
\bea{imA}
{\cal I}m_s\left( {\cal A} \right)&=&\frac{s}{(2\pi)^2}\int\frac{d^2\vec q_1}{\vec
q_1^{\,\, 2}}\Phi_1(\vec q_1,s_0)\int
\frac{d^2\vec q_2}{\vec q_2^{\,\,2}} \Phi_2(-\vec q_2,s_0) \\
&\times& \int\limits^{\delta +i\infty}_{\delta
-i\infty}\frac{d\omega}{2\pi i}\left(\frac{s}{s_0}\right)^\omega
G_\omega (\vec q_1, \vec q_2)\, . \nonumber
\eea
This representation for the amplitude is valid with NLA accuracy.
Here $\Phi_{1}(\vec q_1,s_0)$ and $\Phi_{2}(-\vec q_2,s_0)$
are the impact factors describing the transitions $\gamma^*(p)\to V(p_1)$
and $\gamma^*(p')\to V(p_2)$, respectively.
The Green's function in (\ref{imA}) obeys the BFKL equation
\beq{Green}
\delta^2(\vec q_1-\vec q_2)=\omega \, G_\omega (\vec q_1, \vec q_2)-
\int d^2 \vec q \, K(\vec q_1,\vec q)\, G_\omega (\vec q, \vec q_2) \;,
\eeq
where $K(\vec q_1,\vec q_2)$ is the BFKL kernel.
The scale $s_0$ is artificial. It is introduced in the BFKL approach at the time
to perform the Mellin transform from the $s$-space to the complex angular
momentum plane and must disappear in the full expression for the amplitude
at each fixed order of approximation. Using the result for the meson
NLA impact factor such cancellation was demonstrated explicitly in
Ref.~\cite{IKP04} for the process in question.

\section{The impact factor for the virtual photon to light vector meson transition}

The definition of impact factor (IF) has been given in Ref.~\cite{FF98}; 
in the case of scattering of the particle $A$ off a Reggeized gluon with momentum 
$q_1$, for transverse momentum $\vec \Delta$ and singlet color representation 
in the $t$-channel, the IF has the form~\cite{FFKP99}
\[
\Phi_{A\to A^\prime}(\vec q_1, \vec \Delta, s_0)
= \frac{\delta^{c c'}}{\sqrt{N_c^2-1}} 
\left[\left( \frac{s_0}{\vec q_1^{\:2}} \right)^{\frac{1}{2}
\omega(-\vec q_1^{\:2})}\left( \frac{s_0}{(\vec q_1 - 
\vec \Delta)^2}
\right)^{\frac{1}{2}\omega(-(\vec q_1 - \vec \Delta) ^2)} \right.
\]
\beq{13}
\times \left. \sum_{\{f\}}\int\frac{d\kappa} {2\pi} \theta(s_\Lambda - \kappa)
d\rho_f\Gamma^c_{A\{f\}}\left( \Gamma^{c^\prime}_{A^\prime\{f\}} \right)^*\right]
\eeq
\[
-\frac{1}{2}\!\int\!\!\!\frac{d^{D-2}k}
{\vec k^2(\vec k - \vec \Delta)^2}
\Phi_{A\to A^\prime}^{Born}(\vec k, \vec \Delta, s_0){\cal K}_r^{Born}
(\vec k,\vec q_1, \vec \Delta) \ln\left( \frac{s_\Lambda^2}{s_0(\vec k 
- \vec q_1)^2} \right)\;. 
\]
Here $\omega(t)$ is the Reggeized gluon trajectory in the LLA.
The integration in the first term of Eq.~(\ref{13}) is done 
over the phase space $d\rho_f$ and over the squared invariant mass $\kappa$ of 
the system $\{f\}$ produced in the fragmentation region of the particle $A$, 
$\Gamma^c_{A\{f\}}$ are the related particle-Reggeon effective vertices. 
The second term in Eq.~(\ref{13}) is the counterterm for the LLA part of the 
first one, so that the logarithmic dependence of both terms on the intermediate 
parameter $s_\Lambda \rightarrow \infty$ disappears in their sum. The scale
$s_0$ is artificial and must disappear in the amplitude, to the given accuracy.
The definition~(\ref{13}) guarantees the infrared finiteness of the IFs of 
colorless particles~\cite{FM99}.

\begin{figure}[tb]
\centering
{\epsfysize 4cm \epsffile{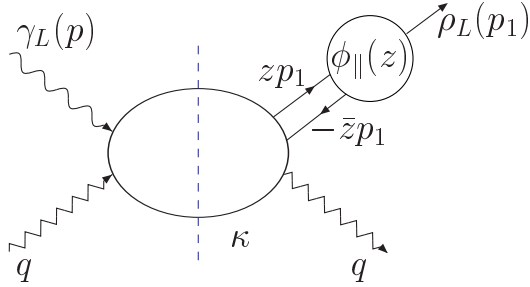}}
\caption[]{The kinematics of the virtual photon to vector meson impact factor.}
\label{fig:impfact}
\end{figure}

Here we study the NLA forward ($\vec \Delta=0$) IF for the transition of 
a virtual photon to a 
light neutral meson $\Phi_{\gamma^*\to V}$, $V=\rho^0, \omega, \phi$
(see Fig.~1). We use the auxiliary Sudakov vectors $p_1$ and $p_2$, such that 
$p_1^2=p_2^2=0$ and $2 (p_1p_2)=s$. The virtual photon momentum is 
$p = p_1-(Q^2/s) p_2$, while Reggeon momenta are
\bea{reggeonM}
q&=&\frac{\kappa +Q^2+\vec q^{\:2}}{s}p_2+q_\perp \, , \quad q^2=q^2_\perp=
-\vec q^{\:2} \, , \nonumber \\
q^\prime&=&\frac{\kappa +\vec q^{\:2}}{s}p_2+q_\perp \; .
\eea
In the forward case under consideration the momentum transfer vector has 
only the longitudinal component, $q-q^\prime =\zeta p_2$. Both the square 
of the Reggeon transverse momentum $\vec q^{\, \, 2}$ and the virtuality 
of the photon $Q^2$ are assumed to be much larger than any hadronic scale. 
Thus we neglect all power suppressed contributions and therefore the mass of
the vector meson mass is put equal to zero and its momentum is
identified with the Sudakov vector $p_1$. 

It is possible to show (see Ref.~\cite{IKP04} for details) that in this kinematics 
the IF can be calculated in the collinear factorization 
framework~\cite{earlyCZ,earlyBL,earlyER} which was developed for the QCD 
description of hard exclusive processes. It turns out that the dominant helicity 
amplitude is a transition of the longitudinally polarized photon $\gamma^*_L$ 
into the longitudinally polarized meson $V_L$, and that both in LLA and in 
NLA the expression for the IF factorizes into the 
convolution\footnote{Here and in the following we consider the {\em color 
unprojected} IF.}
\beq{fact}
\Phi_{\gamma^*_L\to V_L}(\alpha,s_0)=-\frac{4\pi e_q f_V \delta^{cc^\prime}}{N_c Q} 
\int\limits^1_0\, dz\, T_H(z,\alpha,s_0,\mu_F, \mu_R) \, 
\phi_\parallel(z,\mu_F)  
\eeq
of a perturbatively calculable hard-scattering amplitude, $T_H$, and 
a meson twist-2 distribution amplitude, $\phi_\parallel(z)$~\cite{Braun}. 
Here $\mu_F^2\sim Q^2,\vec q^{\:2}$ is a factorization scale 
at which soft and hard physics factorizes according to Eq.~(\ref{fact}). The 
variable $z$ corresponds to the longitudinal momentum fraction carried by the 
quark, for the antiquark the fraction is $\bar z=1-z$. Finally, we introduced 
the ratio $\alpha = \vec q^{\:2}/Q^2$. For the cases of $\rho^0$, $\omega$ and 
$\phi$ meson production, $e_q$ in Eq.~(\ref{fact}) should be 
replaced by $e/\sqrt{2}$, $e/(3\sqrt{2})$ and $-e/3$, respectively. 

We perform calculations with unrenormalized quantities, the bare strong 
coupling constant $\alpha_S$ and the bare meson distribution amplitude 
$\phi_\parallel^{(0)} (z)$. Therefore the NLA expression for the 
hard-scattering amplitude $T_H$ expressed in terms of these quantities
contains both ultraviolet and infrared divergences, appearing as poles 
in the common dimensional regularization parameter $\varepsilon$. 
The ultraviolet divergences disappear after the strong coupling constant 
renormalization. The surviving infrared divergences are only due
to collinear singularities, the soft singularities cancel as usual after  
summing the ``virtual'' and the ``real'' parts of the radiative corrections.
Since IFs should be infrared-finite objects for physical transitions,
it must be possible to absorb the remaining infrared divergences into the 
definition of the nonperturbative distribution amplitude. This is achieved 
by the substitution of the bare distribution amplitude by the renormalized one
(see Ref.~\cite{IKP04} for details).

The calculation of the lowest order contribution to the IF is
straightforward (see Ref.~\cite{IKP04}). For leading power asymptotics, 
the dominant contribution is given by the production of a longitudinally 
polarized meson. The LLA result for the hard-scattering amplitude entering the 
IF is
\beq{LO}
T_H^{(0)}(z,\alpha, s_0,\mu_F ,\mu_R)= \alpha_S\frac{ \alpha}{\alpha +z\bar z} 
\; .
\eeq
Due to the collinear factorization, which effectively puts some fermion lines 
on the mass-shell, the complexity of the intermediate state contributing to 
the IF is reduced in comparison to the case of the virtual photon IF
$\Phi_{\gamma^*\to \gamma^*}$. Actually things go as if we have 
one particle less in the intermediate state.

In the NLA there are two contributions to the IF, from the 
two particle quark-antiquark $(q \bar q)$ and from three-particle
quark-antiquark-gluon $(q\bar q g)$ intermediate states:
\beq{int10}
T_{H}^{(1)}=T^{(q\bar q)}+T^{(q\bar q g)} \, .
\eeq
To calculate the IF in the NLA one has to know the  
$(q \bar q)$ production vertices with NLA accuracy and the $(q\bar q g)$ ones 
at the Born level.   
To calculate $T^{(q\bar q)}$ one needs to convolute the NLA photon-Reggeon 
vertex $\Gamma^{(1)}_{\gamma_L^* q\bar q}$~\cite{FIK} with the Born Reggeon-meson 
vertex $\Gamma^{(0)}_{V_L^* q\bar q}$~\cite{IKP04} and the Born photon-Reggeon 
vertex $\Gamma^{(0)}_{\gamma_L^* q\bar q}$~\cite{IG,FIK} with the NLA Reggeon-meson 
vertex $\Gamma^{(1)}_{V_L^* q\bar q}$~\cite{IKP04}.
To evaluate $T^{(q\bar q g)}$ one needs to convolute Born photon-Reggeon 
vertex $\Gamma^{(0)}_{\gamma_L^* q\bar q g}$~\cite{FIK} with the Born Reggeon-meson 
vertex $\Gamma^{(0)}_{V_L^* q\bar q g}$~\cite{IKP04}.

Taking also into account the so-called $s_\Lambda$-counterterm (last term
in Eq.~(\ref{13})) and summing up all the contributions, one gets for the
{\em renormalized} hard scattering amplitude 
\beq{res}
T_H(z,\alpha,s_0,\mu_F,\mu_R)|_{\alpha\to 0}=
\alpha_S(\mu_R)\frac{\alpha}{\alpha +z\bar z} 
\left\{ 1+\frac{\alpha_S(\mu_R)}{4\pi}\left[\tau(z)+\tau(\bar z)\right]
\right\} \; ,
\eeq
where the expression for $\tau(z)$ is given in Ref.~\cite{IKP04}.

Using this result and the forward BFKL Green's function it is possible
to build the forward amplitude of the process $\gamma_1^*(Q_1^2)\gamma_2^*(Q_2^2)
\to \rho_1 \rho_2$ in the NLA. 

\section{Building up the amplitude}

It is convenient to work in the transverse momentum representation,
where ``transverse'' is related to the plane orthogonal to the vector mesons momenta.
In this representation, defined by
\beq{transv}
\hat{\vec q}\: |\vec q_i\rangle = \vec q_i|\vec q_i\rangle\;,
\eeq
\beq{norm}
\langle\vec q_1|\vec q_2\rangle =\delta^{(2)}(\vec q_1 - \vec q_2) \;, \;\;\;\;\;
\langle A|B\rangle =
\langle A|\vec k\rangle\langle\vec k|B\rangle =
\int d^2k A(\vec k)B(\vec k)\;,
\eeq
the kernel of the operator $\hat K$ is
\beq{kernel-op}
K(\vec q_2, \vec q_1) = \langle\vec q_2| \hat K |\vec q_1\rangle
\eeq
and the equation for the Green's function reads
\beq{Groper}
\hat 1=(\omega-\hat K)\hat G_\omega\;,
\eeq
its solution being
\beq{Groper1}
\hat G_\omega=(\omega-\hat K)^{-1} \, .
\eeq

The kernel is given as an expansion in the strong coupling,
\beq{kern}
\hat K=\bar \alpha_s \hat K^0 + \bar \alpha_s^2 \hat K^1\;,
\eeq
where
\beq{baral}
{\bar \alpha_s}=\frac{\alpha_s N_c}{\pi}
\eeq
and $N_c$ is the number of colors. In Eq.~(\ref{kern}) $\hat K^0$ is the
BFKL kernel in the LLA, $\hat K^1$ represents the NLA correction.

The impact factors are also presented as an expansion in $\alpha_s$
\beq{impE}
\Phi_{1,2}(\vec q)= \alpha_s \,
D_{1,2}\left[C^{(0)}_{1,2}(\vec q^{\,\, 2})+\bar\alpha_s
C^{(1)}_{1,2}(\vec
q^{\,\, 2})\right] \, , \quad D_{1,2}=-\frac{4\pi e_q  f_V}{N_c Q_{1,2}}
\sqrt{N_c^2-1}\, ,
\eeq
where $f_V$ is the meson dimensional coupling constant ($f_{\rho}\approx
200\, \rm{ MeV}$) and $e_q$ should be replaced by $e/\sqrt{2}$, $e/(3\sqrt{2})$
and $-e/3$ for the case of $\rho^0$, $\omega$ and $\phi$ meson production,
respectively.

In the collinear factorization approach the meson transition impact factor
is given as a convolution of the hard scattering amplitude for the
production of a collinear quark--antiquark pair with the meson distribution
amplitude (DA). The integration variable in this convolution is the fraction $z$
of the meson momentum carried by the quark ($\bar z\equiv 1-z$ is
the momentum fraction carried by the antiquark):
\beq{imps1}
C^{(0)}_{1,2}(\vec q^{\,\, 2})=\int\limits^1_0 dz \,
\frac{\vec q^{\,\, 2}}{\vec q^{\,\, 2}+z \bar zQ_{1,2}^2}\phi_\parallel (z)
\, .
\eeq

The NLA correction to the hard scattering amplitude, for a photon with virtuality
equal to $Q^2$, is defined as follows
\beq{imps2}
C^{(1)}(\vec q^{\,\, 2})=\frac{1}{4 N_c}\int\limits^1_0 dz \,
\frac{\vec q^{\,\, 2}}{\vec q^{\,\, 2}+z \bar zQ^2}[\tau(z)+\tau(1-z)]
\phi_\parallel (z)
\, ,
\eeq
with $\tau(z)$ given in the Eq.~(75) of Ref.~\cite{IKP04}.
$C^{(1)}_{1,2}(\vec q^{\,\, 2})$ are given by the previous expression with
$Q^2$ replaced everywhere in the integrand by $Q^2_1$ and $Q^2_2$,
respectively.

The distribution amplitude may be presented as an expansion in Gegenbauer
polynomials
\beq{DA}
\phi_\parallel (z,\mu_F)=6z(1-z)\left[1+a_2(\mu_F)C_2^{3/2}(2z-1)
+ a_4(\mu_F)C_4^{3/2}(2z-1)+\dots\right].
\eeq
The scale dependence of $a_n(\mu_F)$ is well known~\cite{earlyCZ,earlyBL,earlyER}:
\beq{DAsd}
a_n(\mu_F)=L^{\gamma_n/\beta_0}a_n(\mu)\; ,
\eeq
where $L=\alpha_s(\mu_F)/\alpha_s(\mu)$ and
\beq{beta00}
\beta_0=\frac{11 N_c}{3}-\frac{2 n_f}{3}
\eeq
is the leading coefficient of the QCD $\beta$-function, with $n_f$ the number
of active quark flavors. The anomalous dimensions $\gamma_n$ are positive and
grow with $n$. Therefore any DA approaches the asymptotic form
$\phi^{\rm as}_\parallel(z)=6z(1-z)$ at large
$\mu_F$.\footnote{The dependence
of the resulting amplitude on $\mu_F$ is subleading. Due to the collinear
counterterm, see Eq.~(72) of \cite{IKP04}, the NLA correction to the meson impact
factor contains a term proportional to $\ln(\mu_F)$, see Eq.~(75) of \cite{IKP04},
which compensates in the amplitude with NLA accuracy the effect of the
meson DA variation with $\mu_F$.}

Below we will use the DA in the asymptotic form. Besides the simplicity of
the following presentation, the reason is twofold.
Presumably, the
form of DA chosen at low $\mu_F$ will affect mainly only the overall normalization
of the amplitude but not the sum of BFKL energy logarithms and the
resulting dependence of the amplitude on  the  energy in which we
are primarily interested in this study. Another point is that, according to
QCD sum rules estimates~\cite{Braun}, $a_2$(1 GeV) is $0.18\pm 0.10$ for
$\rho$ and $0\pm 0.1$ for $\phi$. Therefore $\phi_\parallel^{\rm as}$ may be indeed a 
good approximation for the DA of light vector mesons.
Integrating over $z$ in~(\ref{imps1})
with $\phi_\parallel(z,\mu_F^2)=\phi^{\rm as}_\parallel(z)$, we obtain,
for photon virtuality $Q^2$,
\beq{al1}
C^{(0)}\,\left(\alpha=\frac{\vec q^{\,\, 2}}{Q^2}\right)\,
=\,6\, \alpha \left[1-\, \frac{\alpha}{c}\,
\ln\frac{2c+1}{2c-1}
\right]\, ,
\eeq
where $c=\sqrt{\alpha+1/4}\,$. $C^{(0)}_{1,2}$ are given by the previous expression
with $Q^2$ replaced by $Q^2_1$ and $Q^2_2$, respectively.
For the NLA term $C^{(1)}_{1,2}(\vec
q^{\:2})$ the integration over $z$ can be performed by a numerical calculation.

To determine the amplitude with NLA accuracy we need an approximate
solution of Eq.~(\ref{Groper1}). With the required accuracy this solution
is
\beq{exp}
\hat G_\omega=(\omega-\bar \alpha_s\hat K^0)^{-1}+
(\omega-\bar \alpha_s\hat K^0)^{-1}\left(\bar \alpha_s^2 \hat K^1\right)
(\omega-\bar \alpha_s \hat
K^0)^{-1}+ {\cal O}\left[\left(\bar \alpha_s^2 \hat K^1\right)^2\right]
\, .
\eeq

The basis of eigenfunctions of the LLA kernel,
\beq{KLLA}
\hat K^0 |\nu\rangle = \chi(\nu)|\nu\rangle \, , \;\;\;\;\;\;\;\;\;\;
\chi (\nu)=
2\psi(1)-\psi\left(\frac{1}{2}+i\nu\right)-\psi\left(\frac{1}{2}-i\nu\right)\, ,
\eeq
is given by the following set of functions:
\beq{nuLLA}
\langle\vec q\, |\nu\rangle =\frac{1}{\pi \sqrt{2}}\left(\vec q^{\,\, 2}\right)
^{i\nu-\frac{1}{2}} \;,
\eeq
for which the  orthonormality  condition takes the form
\beq{ort}
\langle \nu^\prime | \nu\rangle =\int \frac{d^2\vec q}
{2 \pi^2 }\left(\vec q^{\,\, 2}\right)
^{i\nu-i\nu^\prime -1}=\delta(\nu-\nu^\prime)\, .
\eeq
The action of the full NLA BFKL kernel on these functions may be expressed
as follows:
\bea{Konnu}
\hat K|\nu\rangle &=&
\bar \alpha_s(\mu_R) \chi(\nu)|\nu\rangle
 +\bar \alpha_s^2(\mu_R)
\left(\chi^{(1)}(\nu)
+\frac{\beta_0}{4N_c}\chi(\nu)\ln(\mu^2_R)\right)|\nu\rangle
\nonumber \\
&+& \bar
\alpha_s^2(\mu_R)\frac{\beta_0}{4N_c}\chi(\nu)\left(i\frac{\partial}{\partial \nu}
\right)|\nu\rangle \;,
\eea
where the first term represents the action of LLA kernel, while the second
and the third ones stand for the diagonal and the non-diagonal parts of the
NLA kernel.
The function $\chi^{(1)}(\nu)$, calculated
in Ref.~\cite{NLA-kernel}, is conveniently represented in the form
\beq{ch11}
\chi^{(1)}(\nu)=-\frac{\beta_0}{8\, N_c}\left(
\chi^2(\nu)-\frac{10}{3}\chi(\nu)-i\chi^\prime(\nu)
\right) + {\bar \chi}(\nu)\, ,
\eeq
where
\bea{chibar}
\bar \chi(\nu)\,&=&\,-\frac{1}{4}\left[\frac{\pi^2-4}{3}\chi(\nu)-6\zeta(3)-
\chi^{\prime\prime}(\nu)-\frac{\pi^3}{\cosh(\pi\nu)}
\right.
\nonumber \\
&+& \left.
\frac{\pi^2\sinh(\pi\nu)}{2\,\nu\, \cosh^2(\pi\nu)}
\left(
3+\left(1+\frac{n_f}{N_c^3}\right)\frac{11+12\nu^2}{16(1+\nu^2)}
\right)
+\,4\,\phi(\nu)
\right] \, ,
\eea
\beq{phi}
\phi(\nu)\,=\,2\int\limits_0^1dx\,\frac{\cos(\nu\ln(x))}{(1+x)\sqrt{x}}
\left[\frac{\pi^2}{6}-\mbox{Li}_2(x)\right]\, , \;\;\;\;\;
\mbox{Li}_2(x)=-\int\limits_0^xdt\,\frac{\ln(1-t)}{t} \, .
\eeq
Here and below $\chi^\prime(\nu)=d( \chi (\nu) )/d\nu$ and $\chi^{\prime\prime}
(\nu)=d^2(\chi (\nu) )/d^2\nu$.

We will need also the $|\nu\rangle$ representation for the impact factors, which is
defined by the following expressions
\beq{nuu}
\frac{C_1^{(0)}(\vec q^{\,\, 2})}{\vec
q^{\,\, 2}}=\int\limits_{-\infty}^{+\infty}\, d\, \nu^\prime \,c_1(\nu^\prime)
\langle\nu^\prime| \vec q\rangle \;, \;\;\;\;\;
\frac{C_2^{(0)}(\vec q^{\,\, 2})}{\vec q^{\,\, 2}}=\int\limits_{-\infty}^{+\infty}
\, d\, \nu \,c_2(\nu)\,\langle\vec q|\nu\rangle \;,
\eeq
\beq{imp1}
c_1(\nu)=\int d^2\vec q \,\, C_1^{(0)}(\vec q^{\, 2})
\frac{\left(\vec q^{\, 2}\right)^{i\nu-\frac{3}{2}}}{\pi \sqrt{2}}
\, ,\;\;\;\;\;
c_2(\nu)=\int d^2\vec q \,\, C_2^{(0)}(\vec q^{\, 2})
\frac{\left(\vec q^{\, 2}\right)^{-i\nu-\frac{3}{2}}}{\pi \sqrt{2}} \, ,
\eeq
and by similar equations for $c_1^{(1)}(\nu)$ and $c_2^{(1)}(\nu)$
from the NLA corrections to the impact factors, $C_1^{(1)}(\vec
q^{\,\, 2})$ and $C_2^{(1)}(\vec q^{\,\, 2})$.

Using (\ref{exp}) and (\ref{Konnu}) one can derive, after some algebra,
the following representation for the amplitude
\[
\frac{{\cal I}m_s\left( {\cal A} \right)}{D_1D_2}=\frac{s}{(2\pi)^2}
\int\limits^{+\infty}_{-\infty}
d\nu \left(\frac{s}{s_0}\right)^{\bar \alpha_s(\mu_R) \chi(\nu)}
\alpha_s^2(\mu_R) c_1(\nu)c_2(\nu)
\]
\beq{amplnla}
\times \left[1+\bar \alpha_s(\mu_R)
\left(\frac{c^{(1)}_1(\nu)}{c_1(\nu)}+\frac{c^{(1)}_2(\nu)}{c_2(\nu)}\right)
+\bar \alpha_s^2(\mu_R)\ln\left(\frac{s}{s_0}\right)\biggl(\bar\chi(\nu)
\biggr.\right.
\eeq
\[
\left.\left.
+\frac{\beta_0}{8N_c}\chi(\nu)\left[-\chi(\nu)+\frac{10}{3}
+i\frac{d\ln(\frac{c_1(\nu)}{c_2(\nu)})}{d\nu}+2\ln(\mu_R^2)\right]
\right)\right] \; .
\]
We find that
\beq{impsnu}
c_{1,2}(\nu)=
\frac{\left(Q^2_{1,2}\right)^{\pm i\nu-\frac{1}{2}}}{\sqrt{2}}
\frac{\Gamma^2 [\frac{3}{2}\pm i\nu]}{\Gamma [3\pm 2i\nu]}
\frac{6\pi}{\cosh (\pi
\nu)}\, ,
\eeq
\beq{c1c2s}
c_{1}(\nu)c_{2}(\nu)=\frac{1}{Q_1Q_2}\left(\frac{Q_1^2}{Q_2^2}\right)^{i\nu}
\frac{9\,\pi^3(1+4\nu^2)\sinh(\pi\nu)}{32\,\nu\,(1+\nu^2)\cosh^3(\pi\nu)} \, ,
\eeq
\bea{logratio}
i\frac{d\ln(\frac{c_1(\nu)}{c_2(\nu)})}{d\nu}&=&2\biggl[
\psi(3+2i\nu)+\psi(3-2i\nu) \biggr.\\
&-& \left.\psi\left(\frac{3}{2}+i\nu\right)
-\psi\left(\frac{3}{2}-i\nu\right)-\ln\left(Q_1Q_2\right)
\right] \, . \nonumber
\eea

It can be useful to separate from the NLA correction to the impact
factor the terms containing the dependence on $s_0$ and on $\beta_0$,
\beq{sepa}
C^{(1)}(\vec q^{\,\,2})=\int\limits^1_0 dz \,
\frac{\vec q^{\,\, 2}}{\vec q^{\,\, 2}+z \bar zQ^2}\phi_\parallel (z)
\eeq
\[
\times \left[
\frac{1}{4}\ln\left(\frac{s_0}{Q^2}\right)\ln\left(\frac{(\alpha+z\bar
z)^4}{\alpha^2 z^2\bar z^2}\right)+\frac{\beta_0}{4N_c}\left(
\ln\left(\frac{\mu_R^2}{Q^2}\right)+\frac{5}{3}-\ln(\alpha)\right)
+\dots
\right] \;.
\]
Accordingly, one can write
\beq{ffss}
c^{(1)}_{1,2}(\nu)=\tilde c^{(1)}_{1,2}(\nu)+\bar c^{(1)}_{1,2}(\nu) \; ,
\eeq
where $\tilde c^{(1)}_{1,2}(\nu)$ are the contributions from the terms isolated
in the previous equation and $\bar c^{(1)}_{1,2}(\nu)$ represent the rest.
After straightforward calculations we found that
\beq{khk}
\frac{\tilde c^{(1)}_{1}(\nu)}{c_{1}(\nu)}+\frac{\tilde
c^{(1)}_{2}(\nu)}{c_{2}(\nu)}=\ln\left(\frac{s_0}{Q_1Q_2}\right)\chi(\nu)
+\frac{\beta_0}{2N_c}\left[\ln\left(\frac{\mu_R^2}{Q_1Q_2}\right)+\frac{5}{3}\right.
\eeq
\[
+\left.
\psi(3+2i\nu)+\psi(3-2i\nu)-\psi\left(\frac{3}{2}+i\nu\right)
-\psi\left(\frac{3}{2}-i\nu\right) \right] \, .
\]

Using Eq.~(\ref{amplnla}) we construct the following representation for the
amplitude
\bea{series}
\frac{Q_1Q_2}{D_1 D_2} \frac{{\cal I}m_s {\cal A}}{s} &\!=\!\!&
\frac{1}{(2\pi)^2}  \alpha_s(\mu_R)^2 \label{honest_NLA} \\
& \!\times\!\! &
\biggl[ b_0
+\sum_{n=1}^{\infty}\bar \alpha_s(\mu_R)^n   \, b_n \,
\biggl(\ln\left(\frac{s}{s_0}\right)^n \!+
d_n(s_0,\mu_R)\ln\left(\frac{s}{s_0}\right)^{n-1} \biggr)\!
\biggr], \nonumber
\eea
where the coefficients
\beq{bs}
\frac{b_n}{Q_1Q_2}=\int\limits^{+\infty}_{-\infty}d\nu \,  c_1(\nu)c_2(\nu)
\frac{\chi^n(\nu)}{n!} \, ,
\eeq
are determined by the kernel and the impact factors in LLA.
Note that
\beq{b0}
b_0=\frac{9\pi}{4}\left(7\zeta(3)-6\right) \, ,
\eeq
therefore in the Born (the 2-gluon exchange) limit our result coincides with
that of Ref.~\cite{PSW}.

The coefficients
\[
d_n=n\ln\left(\frac{s_0}{Q_1Q_2}\right)+\frac{\beta_0}{4N_c}
\left(
(n+1)\frac{b_{n-1}}{b_n}\ln\left(\frac{\mu_R^2}{Q_1Q_2}\right)
-\frac{n(n-1)}{2} \right.
\]
\beq{ds}
\left.
+ \frac{Q_1Q_2}{b_n}\int\limits^{+\infty}_{-\infty}d\nu \, (n+1)f(\nu)
c_1(\nu)c_2(\nu)
\frac{\chi^{n-1}(\nu)}{(n-1)!}\right)
\eeq
\[
+\frac{Q_1Q_2}{b_n}\left(
\int\limits^{+\infty}_{-\infty}d\nu\, c_1(\nu)c_2(\nu)
\frac{\chi^{n-1}(\nu)}{(n-1)!}\left[
\frac{\bar c^{(1)}_{1}(\nu)}{c_{1}(\nu)}+\frac{\bar
c^{(1)}_{2}(\nu)}{c_{2}(\nu)}
 +(n-1)\frac{\bar \chi(\nu)}{\chi(\nu)}\right]
\right)
\]
are determined by the NLA corrections to the kernel and to the impact
factors. Here we use the notation
\beq{fv}
f(\nu)=\frac{5}{3}+\psi(3+2i\nu)+\psi(3-2i\nu)-\psi\left(\frac{3}{2}+i\nu\right)
-\psi\left(\frac{3}{2}-i\nu\right) \, .
\eeq

One should stress that both representations of the amplitude~(\ref{series})
and~(\ref{amplnla}) are equivalent with NLA accuracy, since they differ only by
 next-to-NLA (NNLA)  terms. 
Actually there exist infinitely many possibilities to write a NLA
amplitude. For instance, another possibility could be to exponentiate the bulk
of the kernel NLA corrections 
\bea{amplnlaE}
\frac{{\cal I}m_s\left( {\cal A} \right)}{D_1D_2}&=&\frac{s}{(2\pi)^2}
\int\limits^{+\infty}_{-\infty}
d\nu \left(\frac{s}{s_0}\right)^{\bar \alpha_s(\mu_R)
\chi(\nu)+\bar \alpha_s^2(\mu_R)
\left(
\bar
\chi(\nu)+\frac{\beta_0}{8N_c}\chi(\nu)\left[-\chi(\nu)+\frac{10}{3}
\right] \right)} \nonumber \\
&\times& \alpha_s^2(\mu_R) c_1(\nu)c_2(\nu)
\left[1+\bar \alpha_s(\mu_R)
\left(\frac{c^{(1)}_1(\nu)}{c_1(\nu)}
+\frac{c^{(1)}_2(\nu)}{c_2(\nu)}\right)\right. \\
&+& \left. \bar \alpha_s^2(\mu_R)\ln\left(\frac{s}{s_0}\right)
\frac{\beta_0}{8N_c}\chi(\nu)\left(
i\frac{d\ln(\frac{c_1(\nu)}{c_2(\nu)})}{d\nu}+2\ln(\mu_R^2)
\right)\right]. \nonumber
\eea
This form of the NLA amplitude was used in Ref.~\cite{KIM1} (see also~\cite{KIM2}), 
without account of the last two terms in the second line of (\ref{amplnlaE}), 
for the analysis of the total $\gamma^*\gamma^*$ cross section.

Since as we will shortly see the NLA corrections are very large, the choice
of the representation for the NLA amplitude becomes practically important.
In the present situation, when an approach to the calculation of the  NNLA
corrections is not developed yet, the series
representation~(\ref{series}) is, in our opinion, a natural choice.
It includes in some sense the minimal amount of NNLA contributions; moreover, its
form is the closest one to the initial goal of the BFKL approach, i.e. to sum
selected contributions in the perturbative series.

It is easily seen from 
 Eqs.~(\ref{series})-(\ref{fv})  that the amplitude is
independent in the NLA from the choice of energy and strong coupling
scales. Indeed, with the required accuracy,
\beq{asev}
\bar \alpha_s(\mu_R)=\bar \alpha_s(\mu_0)\left(
1-\frac{\bar \alpha_s(\mu_0)\beta_0}{4N_c}\ln\left(\frac{\mu_R^2}{\mu_0^2}\right)
\right)
\eeq
and therefore terms $\bar \alpha_s^n\ln^{n-1} s\ln s_0$ and $\bar
\alpha_s^n\ln^{n-1} s\ln \mu_R$ cancel in~(\ref{series}).

One can trace the contributions to each $d_n$ coefficient coming from the
NLA corrections to the BFKL kernel and from the NLA impact factors
\beq{dsdec}
d_n=d_n^{\rm{ker}}+d_n^{\rm{IF}}\, ,
\eeq
\[
d_n^{\rm{IF}}=n\ln\left(\frac{s_0}{Q_1Q_2}\right)
\]
\beq{dsim}
+\frac{\beta_0}{4N_c}
2\left(
\frac{b_{n-1}}{b_n}\ln\left(\frac{\mu_R^2}{Q_1Q_2}\right)
+ \frac{Q_1Q_2}{b_n}\int\limits^{+\infty}_{-\infty}d\nu \, f(\nu)
c_1(\nu)c_2(\nu)
\frac{\chi^{n-1}(\nu)}{(n-1)!}\right)
\eeq
\[
+\frac{Q_1Q_2}{b_n}\left(
\int\limits^{+\infty}_{-\infty}d\nu\, c_1(\nu)c_2(\nu)
\frac{\chi^{n-1}(\nu)}{(n-1)!}\left[
\frac{\bar c^{(1)}_{1}(\nu)}{c_{1}(\nu)}+\frac{\bar
c^{(1)}_{2}(\nu)}{c_{2}(\nu)}
\right] \right) \, .
\]
The first coefficient, $d_1$, is entirely due to the NLA corrections to the
impact factors,
\beq{d1}
d_1=d_1^{\rm{IF}}\, , \quad d_1^{\rm{ker}}=0 \,.
\eeq
Let us note that in the BFKL formalism the NLA contribution to the impact
factors guarantees not only independence of the amplitude from the energy
scale, $s_0$, but it contains also a term proportional to $\ln \mu_R$ which is
important for the renorm-invariance of the predicted results, i.e.
the dependence of the amplitude on $\mu_R$ and $s_0$ is subleading to the NLA
accuracy.

\section{Numerical results}

In this Section we present some numerical results for the amplitude given
in Eq.~(\ref{series}) for the $Q_1=Q_2\equiv Q$ kinematics, i.e. in the
``pure'' BFKL regime. The other interesting regime, $Q_1\gg Q_2$ or vice-versa,
where collinear effects could come heavily into the game, will not be 
 considered
here. We will emphasize in particular the dependence on 
the renormalization scale
$\mu_R$ and $s_0$ in the NLA result.

In all the forthcoming figures the quantity on the vertical axis is the
L.H.S. of Eq.~(\ref{series}), ${\cal I}m_s ({\cal A})Q^2/(s \, D_1 D_2)$.
In the numerical analysis presented below we truncate the series in the
R.H.S. of Eq.~(\ref{series}) to $n=20$, after having verified that
this procedure gives a very good approximation of the infinite sum
for the $Y$ values $Y\leq 10$. We use the two--loop running coupling
corresponding to the value $\alpha_s(M_Z)=0.12$.

We have calculated numerically the $b_n$ and $d_n$ coefficients for
$n_f=5$ and $s_0=Q^2=\mu_R^2$, getting
\beq{coe}
\begin{array}{llll}
b_0=17.0664 & & \\ 
b_1=34.5920 & b_2=40.7609 & b_3=33.0618 & b_4=20.7467  \\
b_5=10.5698 & b_6=4.54792 & b_7=1.69128 & b_8=0.554475 \\
& & & \\
d_1=-3.71087 & d_2=-11.3057 & d_3=-23.3879 & d_4=-39.1123 \\
d_5=-59.207 & d_6=-83.0365 & d_7=-111.151 & d_8=-143.06 \;. \\
\end{array}
\eeq
In this case contributions to the $d_n$ coefficients originating from
the NLA corrections to the impact factors are
\beq{coeffim}
\begin{array}{llll}
d_1^{\rm{IF}}=-3.71087 & d_2^{\rm{IF}}=-8.4361 & d_3^{\rm{IF}}=-13.1984 &
d_4^{\rm{IF}}=-18.0971 \\
d_5^{\rm{IF}}=-23.0235 & d_6^{\rm{IF}}=-27.9877 &
d_7^{\rm{IF}}=-32.9676 & d_8^{\rm{IF}}=-37.9618 \;. \\
\end{array}
\eeq
Thus, comparing~(\ref{coe}) and (\ref{coeffim}), we see that the contribution
from the kernel starts to be larger than the impact factor one only for $n\geq 4$.

These numbers make visible the effect of the NLA corrections: the $d_n$
coefficients are negative and increasingly large in absolute values as the
perturbative order increases. The NLA corrections turn to be very large.
In this situation the optimization of perturbative expansion, in our case the
choice of the renormalization scale $\mu_R$ and of the energy scale $s_0$,
becomes an important issue.
Below we will adopt the principle of minimal sensitivity (PMS)~\cite{Stevenson}.
Usually PMS is used to fix the value of the renormalization scale for the strong
coupling. We suggest to use this principle in a broader sense,  requiring 
in our case  the minimal sensitivity of the predictions to the change 
of both the renormalization and the energy scales, $\mu_R$ and $s_0$.

Since the dependence of results on $s_0$ is a feature typical of the BFKL
approach and is somewhat new for the application of PMS, we will first
illustrate the success of PMS in this respect on the following QED result
known since a long time.
In 1937 Racah calculated the total cross section for the production of
$e^+e^-$ pairs in the collisions of two heavy ions at high energies~\cite{Racah},
\beq{HIs}
\sigma=\frac{28\alpha_{EM}^4Z_1^2Z_2^2}{27\pi m_e^2}\left(l^3+A l^2+ B l + C \right)
+{\cal O}\left(\frac{1}{(p_1p_2)}\right)\;;
\eeq
here $Z_{1,2}$ are the ions charges, $m_e$ is electron mass, the ions' four-momenta
are $p_{1,2}$,
\beq{lls}
l=\ln\frac{2(p_1p_2)}{m_1m_2}\, ,
\eeq
is the energy logarithm and $m_{1,2}$ are the masses of the ions. The
contributions suppressed by the power of energy are denoted as
${\cal O}(1/(p_1p_2))$. 

\begin{figure}[tb]
\centering
{\epsfysize 5cm \epsffile{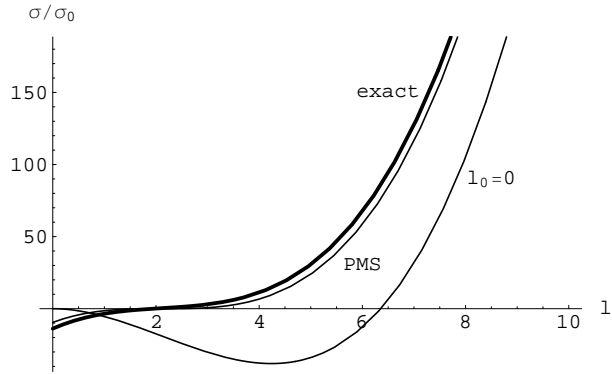}}
\caption[]{$\sigma/\sigma_0$ as a function of the energy logarithm $l$
for the cases
of exact result of Racah, approximated result with $l_0=-A/3$ (PMS
optimal choice) and $l_0=0$ (kinematical scale for energy logarithms).}
\label{HIe}
\end{figure}

The coefficients in front of the subleading logarithms are large and have
alternating signs
\[
A = -178/28=-6.35714
\]
\beq{Rcoef}
B = \frac{1}{28}(7\pi^2 + 370)=15.6817
\eeq
\[
C = -\frac{1}{28}\left(348 + \frac{13}{2}\pi^2 - 21\zeta(3)\right)=-13.8182\, .
\]
To illustrate the application of PMS, imagine that we know only the coefficient $A$
in front of the first subleading logarithm. Then using this knowledge
we can construct the following approximation
\beq{HIapp}
\sigma^{app}=\sigma_0\left((l-l_0)^3+(A+3l_0) (l-l_0)^2
\right)\, , \quad \sigma_0=\frac{28\alpha_{EM}^4Z_1^2Z_2^2}{27\pi m_e^2}\, ,
\eeq
(an analog of NLA in the BFKL approach) where we shift the energy scale
introducing the parameter $l_0$. Note that the dependence of the cross section
on $l_0$ is subleading in the approximation used in Eq.~(\ref{HIapp}).
We fix $l_0$ by requiring the minimal sensitivity of (\ref{HIapp}) to the change of
this parameter. It is not difficult to find that this procedure gives
$l_0=-A/3=2.11905$.\footnote{Note that in this example PMS gives the value
of the parameter $l_0$ for which the correction to the lowest approximation,
$(l-l_0)^3$, vanishes. Therefore in this case PMS gives a result which
coincides with the one given by another alternative approach to optimize
the approximation, the fast apparent convergence prescription~\cite{Grun}.}
In Fig.~\ref{HIe} we present three curves for $\sigma/\sigma_0$ as a function of
the energy logarithm $l$; the first one was calculated using the exact result of
Racah (with all subleading logarithms), the other two curves were calculated
using~(\ref{HIapp}) with $l_0=0$ and with the PMS value $l_0=2.11905$.

We see that the PMS approach gives a very good approximation to the Racah
result\footnote{The negative cross section at $l<2$ is due to the fact that
terms subleading in energy, ${\cal O}(1/(p_1p_2))$ in~(\ref{HIs}),
are not taken into account.}.
On the other hand the procedure with $l_0=0$, which means that a kinematical
scale for energy logarithms is used in the approximate formula, makes an awfully
bad job for the whole $l$ range presented in the figure.

Returning to our problem, we apply PMS to our case requiring the minimal
sensitivity of the amplitude~(\ref{series}) to the variation of $\mu_R$ and
$s_0$. More precisely, we replace in~(\ref{series}) $\ln(s/s_0)$ with $Y-Y_0$,
where $Y=\ln(s/Q^2)$ and $Y_0=\ln(s_0/Q^2)$, and study the dependence of the
amplitude on $Y_0$.

\begin{figure}[tb]
\centering
{\epsfysize 3.7cm \epsffile{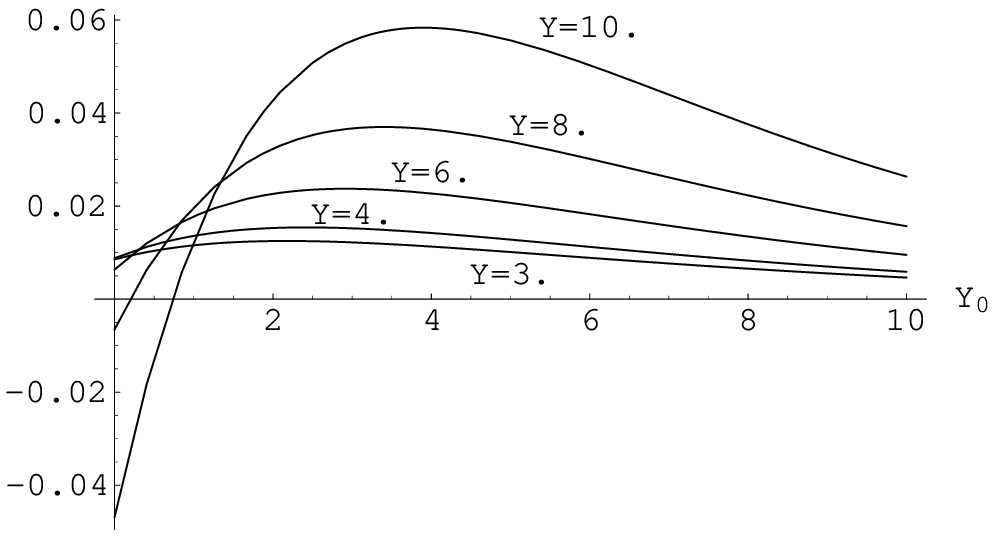}}
{\epsfysize 3.7cm \epsffile{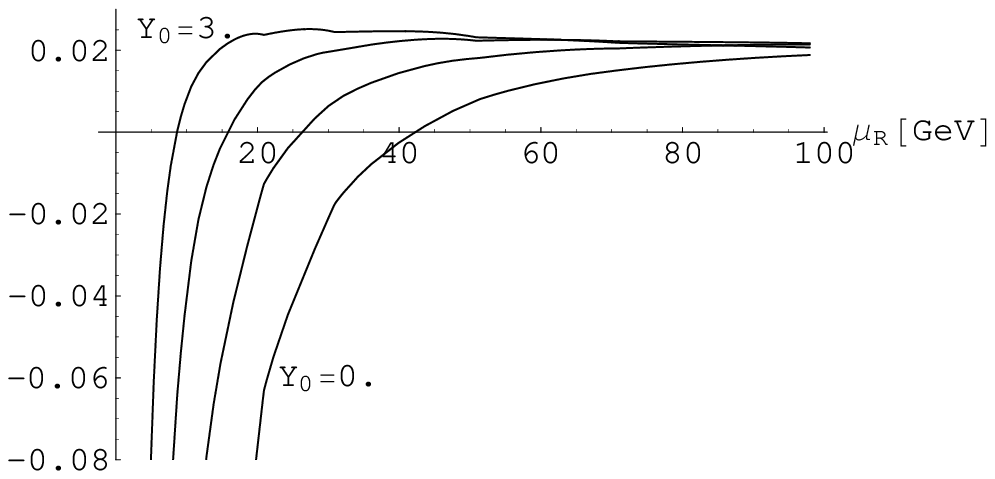}}
\caption[]{(Left) ${\cal I}m_s ({\cal A})Q^2/(s \, D_1 D_2)$ as a function of
$Y_0$ at $\mu_R=10 Q$. The different curves are for $Y$ values of 10, 8, 6, 4 and 3.
The photon virtuality $Q^2$ has been fixed to 24 GeV$^2$ ($n_f=5$).
(Right) ${\cal I}m_s ({\cal A})Q^2/(s \, D_1 D_2)$ as a function of
$\mu_R$ at $Y$=6. The different curves are, from above to below, for $Y_0$
values of 3, 2, 1 and 0. The photon virtuality $Q^2$ has been fixed to
24 GeV$^2$ ($n_f=5$).}
\label{Y0_muR}
\end{figure}

The next two figures illustrate the dependence on these parameters for
$Q^2$=24 GeV$^2$ and $n_f=5$.
In Fig.~\ref{Y0_muR}(left) we show the dependence of amplitude on $Y_0$ for
$\mu_R=10 Q$, when $Y$ takes the values 10, 8, 6, 4, 3.

We see that for each $Y$ the amplitude has an extremum in $Y_0$ near which
it is not sensitive to the variation of $Y_0$, or $s_0$. Our choice of $\mu_R$
for this figure is motivated by the study of $\mu_R$ dependence. In
Fig.~\ref{Y0_muR}(right) we present the $\mu_R$ dependence for $Y=6$; the curves from above
to below are for $Y_0$=3, 2, 1, 0.

\begin{figure}[tb]
\centering
{\epsfysize 5cm \epsffile{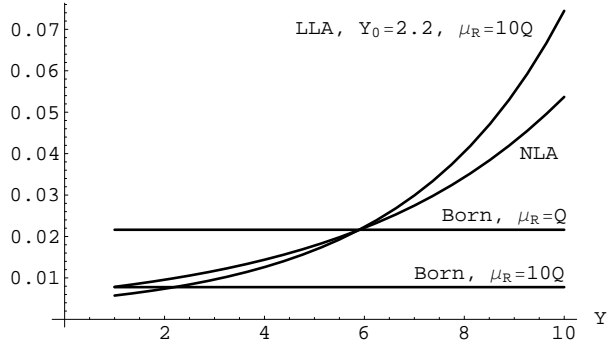}}
\caption[]{${\cal I}m_s ({\cal A})Q^2/(s \, D_1 D_2)$ as a function of
$Y$ for optimal choice of the energy parameters $Y_0$ and $\mu_R$ (curve 
labeled by ``NLA''). 
The other curves represent the LLA result for $Y_0=2.2$ and $\mu_R=10Q$
and the Born (2-gluon exchange) limit for $\mu_R=Q$ and $\mu_R=10Q$.
The photon virtuality $Q^2$ has been fixed to 24 GeV$^2$ ($n_f=5$).}
\label{PMSres}
\end{figure}

Varying $\mu_R$ and $Y_0$ we found for each $Y$ quite large regions in
$\mu_R$ and $Y_0$ where the amplitude is practically independent on $\mu_R$ and
$Y_0$. We use this value as the NLA result for the amplitude at given $Y$.
In Fig.~\ref{PMSres} we present the amplitude found in this way as a function of
$Y$. The resulting curve is compared with the curve obtained from the LLA
prediction when the scales are chosen as $\mu_R=10 Q$ and $Y_0=2.2$, in
order to make the LLA curve the closest possible (of course it is not an exact
statement) to the NLA one in the given interval of $Y$. The two
horizontal lines in Fig.~\ref{PMSres} are the Born (2-gluon exchange)
predictions calculated for $\mu_R=Q$ and $\mu_R=10 Q$.

Similar procedure was applied to a lower value of the photon virtuality,
$Q^2$=5 GeV$^2$ and $n_f=4$ (see Ref.~\cite{IP06} for details).

We stress that one should take with care BFKL predictions for small values
of $Y$, since in this region the contributions suppressed by powers of
the energy should be taken into account. At the lowest order in $\alpha_s$ such
contributions are given by diagrams with quark exchange in the $t$-channel and are
proportional in our case to $\alpha_{EM}\alpha_sf_V^2/Q^2$.
At higher orders power suppressed contributions contain double
logarithms, terms $\sim \alpha_s^n\ln^{2n} s$, which can lead to a significant
enhancement. Such contributions were
recently studied for the total cross section of $\gamma^*\gamma^*$
interactions~\cite{Bartels}.

If the NLA (and LLA) curves in Fig.~\ref{PMSres} are compared
with the Born (2-gluon exchange) results, one can conclude that the summation of
BFKL series gives negative contribution to the Born result for $Y<6$ if one chooses
for the scale of the strong coupling in the Born amplitude the value given by the
kinematics, $\mu_R=Q$. We believe that our calculations show that one should at
least accept with some caution the results obtained in the Born approximation,
since they do not give necessarily an estimate of 
the observable from below.

Another important lesson from our calculation is the very large scale for
$\alpha_s$ (and therefore the small $\alpha_s$ itself) we obtain using PMS.
It appears to be much bigger than the kinematical scale and looks unnatural
since there is no other scale for transverse momenta in the problem at
question except $Q$. Moreover one can guess that at higher orders
the typical transverse momenta are even smaller than $Q$ since they "are shared"
in the many-loop integrals and the strong coupling grows in the infrared.
In our opinion  the  large values of $\mu_R$ we found is not an indication of the
appearance of a new scale, but is rather a manifestation of the nature of the
BFKL series. The fact is that NLA corrections are large and then, necessarily,
since the exact amplitude should be renorm- and energy scale invariant, the
NNLA terms should be large and of the opposite sign with respect to the NLA.
We guess that if the NNLA corrections were known and we would apply PMS to the
amplitude constructed as LLA + NLA-corrections + NNLA-corrections,
we would obtain in such calculation more natural values of $\mu_R$.

In the last years strong efforts have been devoted to the
improvement of the NLA BFKL kernel as a consequence of the analysis of
collinear singularities of the NLA corrections and by the account of
further collinear terms beyond NLA~\cite{Sal98,collinear}. 
This strategy has something in common with ours, in the sense that it is
also inspired by renormalization-group invariance and it also leads to
the addition of terms beyond the NLA. These extra-terms are large and of
opposite sign with respect to the NLA contribution, so that they partially
compensate the NLA corrections. The findings of the present work
suggest, however, that the corrections to the impact factors heavily 
contribute to the NLA amplitude, being even dominating in some interval
of non-asymptotically high energies. 
Moreover, by inspection of the structure of the amplitude in the regime of 
strongly asymmetric photon virtualities, one can deduce that also the
impact factors generate collinear terms which add up to those arising
from the kernel, see e.g. Eqs.~(84) and~(85) of Ref.~\cite{IKP04}. 
This leads us to the conclusion that in the approaches based on kernel 
improvement the additional information coming from impact factors should somehow be
taken into account when available. These issues 
certainly deserve further investigation and we believe that useful hints in this 
direction can be gained from the study of the $\gamma^* \gamma^* \to V V$ amplitude 
in the regime of strongly ordered photon virtualities~\cite{CPS07,Cap08}.

We conclude this Section with a comment on the possible implications of
our results for mesons electroproduction to the phenomenologically 
more important case of the $\gamma^* \gamma^*$ total cross section. 
By numerical inspection we have found that the ratios $b_n/b_0$ we got for the 
meson case agree for $n= 1\div 10$ at $1\div 2\%$ accuracy level with the analogous 
ratios for the longitudinal photon case and at $3.5\div 30\%$ accuracy level 
with those for the transverse photon case. Should this similar behavior persist 
also in the NLA, our predictions could be easily translated to estimates of the
$\gamma^* \gamma^*$ total cross section. 

\section{Study of systematic effects}

It is important to have an estimate
of the systematic uncertainty which plagues our determination of the
energy behavior of the amplitude. The main sources of systematic effects
are given by the choice of the representation of the amplitude and by
the optimization method adopted. In the following, we compare the
determination of the amplitude at $Q^2$ = 24 GeV$^2$ ($n_f=5$)
through the PMS method, given in Fig.~\ref{PMSres}, with other determinations
obtained changing either the representation of the amplitude or the
optimization method.

\begin{figure}[tb]
\centering
{\epsfysize 3.7cm \epsffile{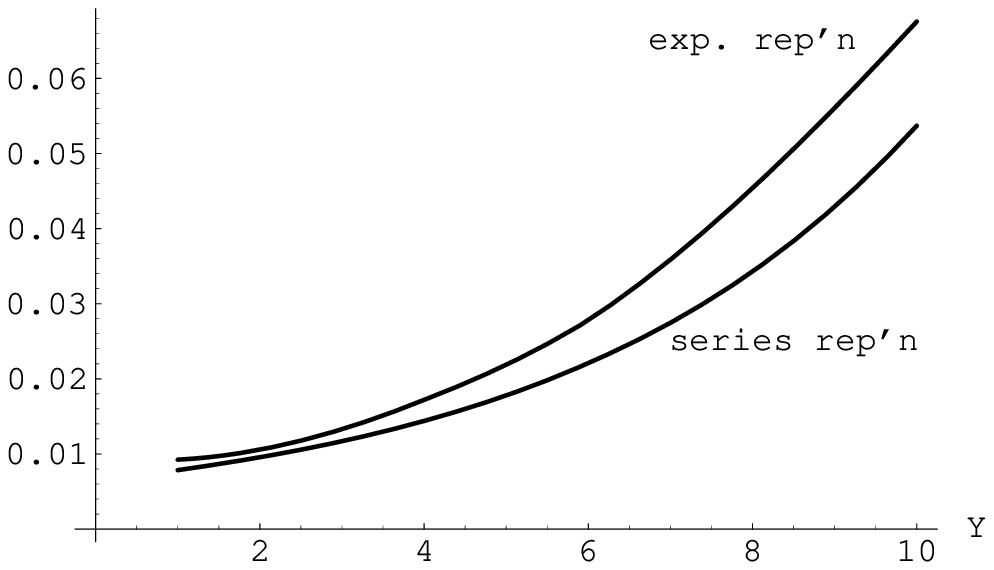}}
{\epsfysize 3.7cm \epsffile{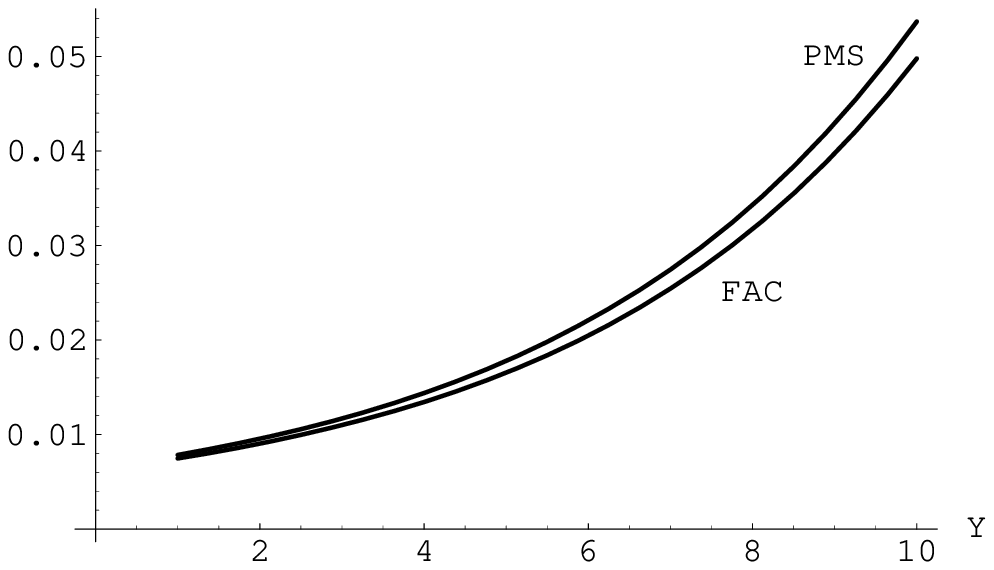}}
\caption[]{(Left) ${\cal I}m_s ({\cal A})Q^2/(s \, D_1 D_2)$ as a function of $Y$
at $Q^2$=24 GeV$^2$ ($n_f=5$) from series and ``exponentiated'' representations,
in both cases with the PMS optimization method. 
(Right) ${\cal I}m_s ({\cal A}_{\mathrm series})Q^2/(s \, D_1 D_2)$ as a function of
$Y$ at $Q^2$=24 GeV$^2$ ($n_f=5$) from the series representation with PMS
and FAC optimization methods.}
\label{PMSexp_PMSseries+FACseries_PMSseries}
\end{figure}

At first, we compare the series and the ``exponentiated'' determinations
using in both case the PMS method. The procedure we followed to determine
the energy behavior of the ``exponentiated'' amplitude is straightforward:
for each fixed value of $Y$ we determined the optimal choice of the
parameters $\mu_R$ and $Y_0$ for which the amplitude given in
Eq.~(\ref{amplnlaE}) is the least sensitive to their variation. Also
in this case we could see wide regions of stability of the amplitude
in the $(\mu_R,Y_0)$ plane. The optimal values of $\mu_R$ and $Y_0$
are quite similar to those obtained in the case of the series representation,
with only a slight decrease of the optimal $\mu_R$. In
Fig.~\ref{PMSexp_PMSseries+FACseries_PMSseries}(left) we show the result and compare it to
the PMS determination from the series representation. The two curves are
in good agreement at the lower energies, the deviation increasing for large
values of $Y$. It should be stressed, however, that the applicability
domain of the BFKL approach is determined by the condition
$\bar \alpha_s(\mu_R) Y \sim 1$ and, for $Q^2$=24 GeV$^2$ and for the
typical optimal values of $\mu_R$, one gets from this condition $Y\sim 5$.
Around this value the discrepancy between the two determinations is within a
few percent.

As a second check, we changed the optimization method and applied
it both to the series and to the ``exponentiated'' representation.
The method considered is the fast apparent convergence (FAC)
method~\cite{Grun}, whose strategy, when applied to a usual perturbative
expansion, is to fix the renormalization scale to the value for which the
highest order correction term is exactly zero. In our case, the application
of the FAC method requires an adaptation, for two reasons: the first is that
we have two energy parameters in the game, $\mu_R$ and $Y_0$, the second
is that, if only strict NLA corrections are taken, the amplitude
does not depend at all on these parameters.

Therefore, in the case of the series representation, Eq.~(\ref{series}), we
choose to put to zero the sum
\[
\frac{1}{(2\pi)^2}  \alpha_s(\mu_R)^2
\sum_{n=1}^{\infty}\bar \alpha_s(\mu_R)^n   \, b_n \,
d_n(s_0,\mu_R)\ln\left(\frac{s}{s_0}\right)^{n-1}
\]
and found for each fixed $Y$ the values of $\mu_R$ and $Y_0$ for which
the vanishing occurs. This gives a line of values in the $(\mu_R,Y_0)$ plane, among
which the optimal choice is done applying a minimum sensitivity criterion.
The result is shown in Fig.~\ref{PMSexp_PMSseries+FACseries_PMSseries}(right).
The agreement with the series representation with the PMS method is rather good over 
a wide energy range.

\begin{figure}[tb]
\centering
{\epsfysize 3.7cm \epsffile{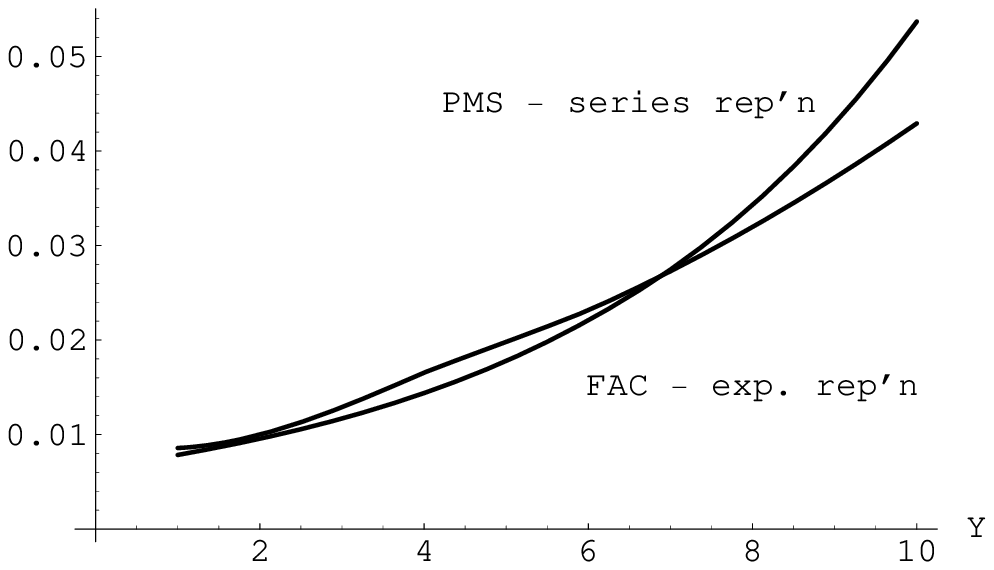}}
{\epsfysize 3.7cm \epsffile{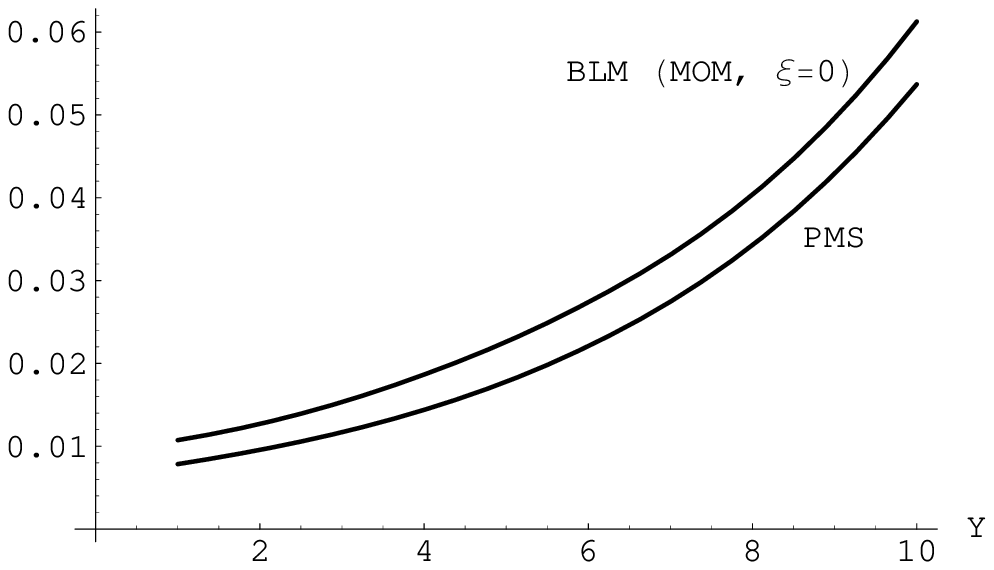}}
\caption[]{(Left) ${\cal I}m_s ({\cal A})Q^2/(s \, D_1 D_2)$ as a function of
$Y$ at $Q^2$=24 GeV$^2$ ($n_f=5$) from the series representation with the PMS
optimization method and from the ``exponentiated'' representation with the FAC
optimization method.
(Right) ${\cal I}m_s ({\cal A})Q^2/(s \, D_1 D_2)$ as a function of
$Y$ at $Q^2$=24 GeV$^2$ ($n_f=5$) from the series representation with PMS
and BLM optimization methods.}
\label{FACexp_vs_PMSseries+BLM_vs_PMS_series}
\end{figure}

In the case of the ``exponentiated'' amplitude'', Eq.~(\ref{amplnlaE}),
we proceeded in the same way, but requiring the vanishing of the
expression given by the R.H.S. of Eq.~(\ref{amplnlaE}) minus
the LLA amplitude, i.e.
\[
\frac{{\cal I}m_s\left( {\cal A}_{\mathrm exp} \right)}{D_1D_2}
-\frac{s}{(2\pi)^2}
\int\limits^{+\infty}_{-\infty}
d\nu \left(\frac{s}{s_0}\right)^{\bar \alpha_s(\mu_R) \chi(\nu)}
\alpha_s^2(\mu_R) c_1(\nu)c_2(\nu)\;.
\]
In Fig.~\ref{FACexp_vs_PMSseries+BLM_vs_PMS_series}(left) the result is compared with 
series representation in the PMS method: there is nice agreement over the
whole energy range considered.

Another popular optimization procedure is the Brodsky-Lepage-Mac\-ken\-zie (BLM)
method~\cite{BLM}, which amounts to perform a finite renormalization
to a physical scheme and then choose the renormalization scale in order to
remove the $\beta_0$-dependent part. We applied this method only to the
series representation, Eq.~(\ref{series}), and proceeded as follows: we
first performed a finite renormalization to the momentum (MOM) scheme
with $\xi=0$ (see Ref.~\cite{KIM1}),
\[
\alpha_s \to \alpha_s \left[1+T_{MOM}(\xi=0) \frac{\alpha_s}{\pi}\right]\;,
\hspace{1cm}
T_{MOM}(\xi=0)=T_{MOM}^{conf}+T_{MOM}^\beta \;,
\]
\[
T_{MOM}^{conf}=\frac{N_c}{8}\frac{17}{2} I \;, \hspace{1cm}
T_{MOM}^\beta=-\frac{\beta_0}{2}\left[1+\frac{2}{3}I\right]\;,
\hspace{1cm} I\simeq 2.3439 \;,
\]
then, we chose $Y_0$ and $\mu_R$ in order to make the term proportional
to $\beta_0$ in the resulting amplitude vanish. We observe that the
$\beta_0$-dependence in the series representation of the amplitude is hidden
into the $d_n$ coefficients, Eq.~(\ref{ds}). Among the resulting pairs of
values for $Y_0$ and $\mu_R$, we determined the optimal one according to
minimum sensitivity. This method has a drawback in our case, since for
each fixed $Y$, the optimal choice for $Y_0$ turned to be always
$Y_0\simeq Y$. However, if one blindly applies the procedure above, one
gets a curve which slightly overshoots the one for the series
representation with the PMS method, see 
Fig.~\ref{FACexp_vs_PMSseries+BLM_vs_PMS_series}(right).

\begin{figure}[tb]
\centering
{\epsfysize 3.7cm \epsffile{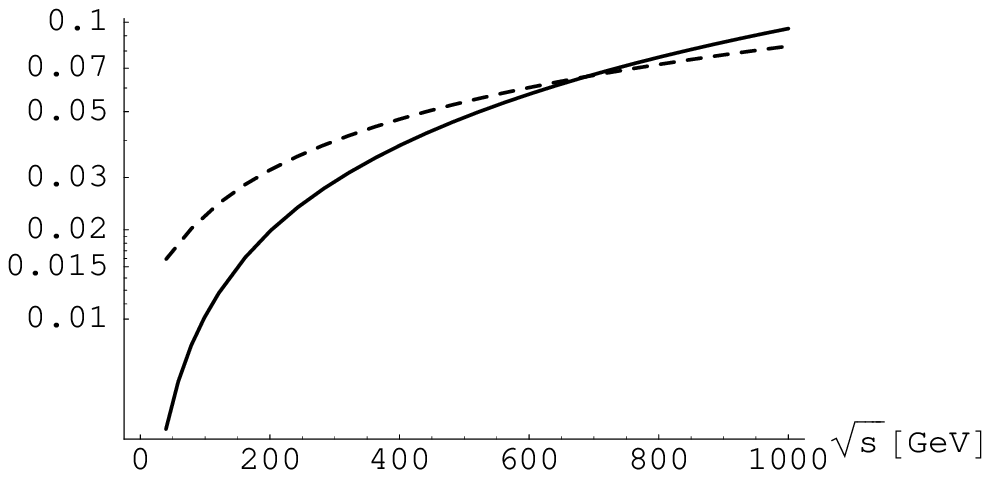}}
{\epsfysize 3.7cm \epsffile{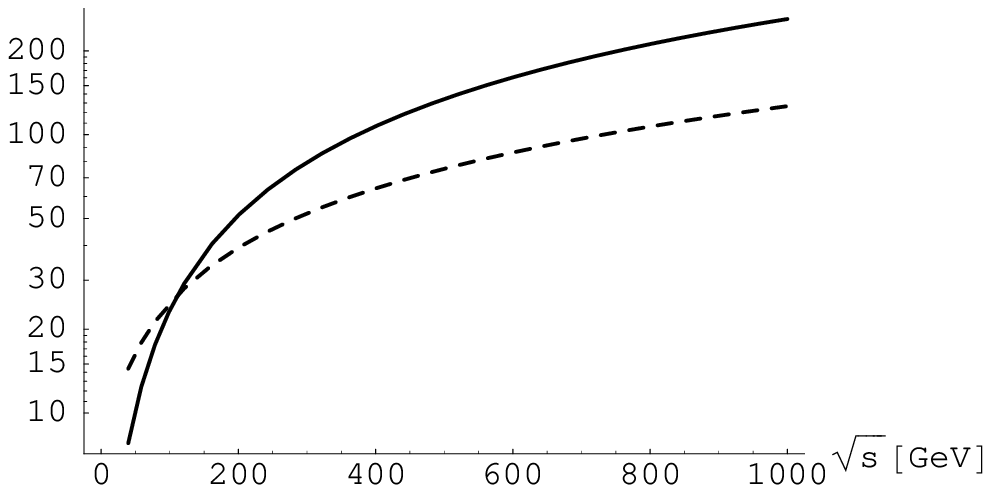}}
\caption[]{(Left) Linear-log plot of $\left. d\sigma/dt\right|_{t=t_0}$ [pb/GeV$^2$] as a function of
$\sqrt{s}$ at $Q^2$=16 GeV$^2$ ($n_f=4$) from the series representation with the PMS
optimization method (solid line) compared with the determination from the approach in
Ref.~\cite{PSW1} (dashed line).
(Right) The same as (left) at $Q^2$=4 GeV$^2$ ($n_f=3$)}
\label{dsig}
\end{figure}

\section{The differential cross section at the minimum $|t|$:
comparison with an approach based on collinear improvement}

The $\gamma^* \gamma^* \to \rho \rho $ process at the lowest order
(two-gluon exchange in the $t$-channel) was studied in Ref.~\cite{PSW}.
At that level our results coincide, see also~\cite{IP06}.
The same process with the inclusion of NLA BFKL effects has been considered
in Ref.~\cite{PSW1}. In that paper,
the amplitude has been built with the following ingredients: leading-order
impact factors for the $\gamma^* \to \rho$ transition, BLM scale fixing
for the running of the coupling in the prefactor of the amplitude (the BLM
scale is found using the NLA $\gamma^* \to \rho$ impact factor
calculated in Ref.~\cite{IKP04}) and renormalization-group-resummed
BFKL kernel, with resummation performed on the LLA BFKL kernel at fixed
coupling~\cite{KMRS04}. In Ref.~\cite{PSW1} the behavior of $d\sigma/dt$ at $t=t_0$
was determined as a function of $\sqrt{s}$ for three values of
the common photon virtuality, $Q$=2, 3 and 4 GeV.

In order to make a comparison with the findings of Ref.~\cite{PSW1},
we computed $d\sigma/dt$ at $t=t_0$ for $Q$=2 and $Q$=4 GeV as functions
of $\sqrt{s}$. We used $f_\rho$=216 MeV, $\alpha_{\mathrm{EM}}=1/137$ and
the two--loop running strong coupling corresponding to the value
$\alpha_s(M_Z)=0.12$. The results are shown in the linear-log
plots of Figs.~\ref{dsig}, which show disagreement. 
This is not surprising in consideration of the approximations adopted 
in Ref.~\cite{PSW1}, 

It would be interesting to understand to what extent
this disagreement is due to the use in Ref.~\cite{PSW1} of
LLA impact factors instead of the NLA ones or to the way the collinear
improvement of the kernel is performed.

\begin{figure}[tb]
\centering
{\epsfysize 3.7cm \epsffile{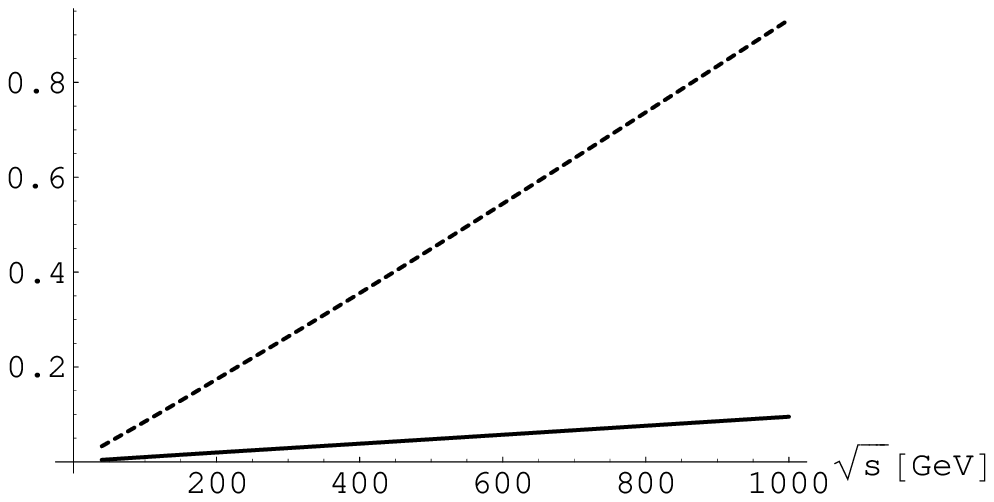}}
{\epsfysize 3.7cm \epsffile{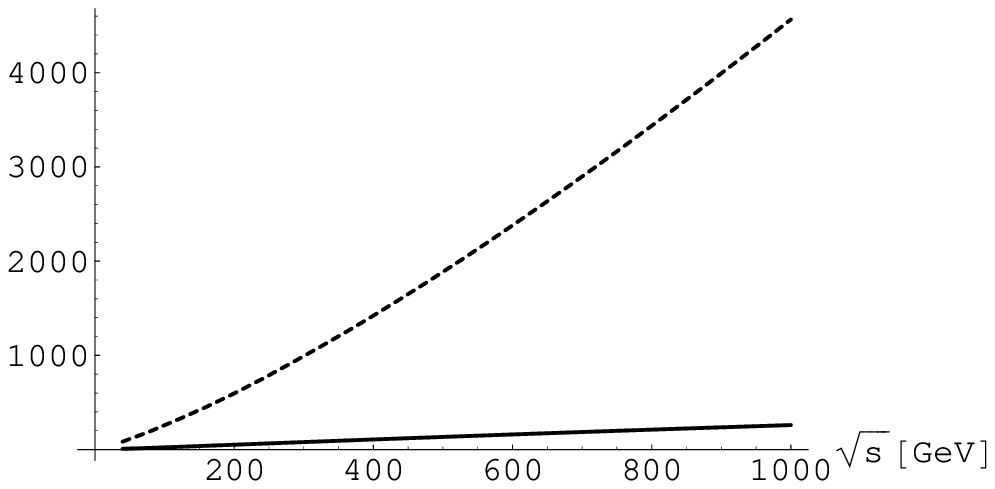}}
\caption[]{(Left) Linear plot of $\left. d\sigma/dt\right|_{t=t_0}$ [pb/GeV$^2$] as a function of
$\sqrt{s}$ at $Q^2$=16 GeV$^2$ ($n_f=4$) from the series representation with the PMS
optimization method using NLO impact factors (solid line) and LO impact factors (dashed line).
(Right) The same as (left) at $Q^2$=4 GeV$^2$ ($n_f=3$).}
\label{dsig_LOvsNLO}
\end{figure}

In order to understand to what extent the discrepancy is due to the use of leading order (LO)
impact factors instead of next-to-leading order (NLO) ones, we repeated our determination of
$d\sigma/dt$ at $t=t_0$ for $Q$=2 and $Q$=4 GeV, using LO impact factors and keeping
from the their NLO contribution only the terms proportional to $\ln[s_0/(Q_1 Q_2)]$
and to $\ln[\mu_R^2/(Q_1 Q_2)]$ which are universal and needed to guarantee the $s_0$-
and $\mu_R$-independence of the amplitude with NLA accuracy. The result is that
$d\sigma/dt$ at $t=t_0$ increases roughly by an order of magnitude with respect to
our previous determination (see Figs.~\ref{dsig_LOvsNLO})
and therefore the disagreement with~\cite{PSW1} becomes even worse. This is not
surprising: impact factors give a sizable contribution to the NLA part of the
amplitude which is negative with respect to the LLA part; if they are kept at LO,
the NLA part of the amplitude is less negative and the total amplitude is therefore
increased.

\section{Conclusions}

We have determined the amplitude for the forward transition from two virtual
photons to two light vector mesons in the Regge limit of QCD with next-to-leading
order accuracy. This amplitude is the first one ever written in the next-to-leading
approximation for a collision process between strongly interacting colorless
particles. It is given as an integral over the $\nu$ parameter, which labels the
eigenvalues of the leading order forward BFKL kernel in the singlet color
representation. This form is suitable for numerical evaluations.
The result obtained is independent on the energy scale $s_0$, and on the
renormalization scale $\mu_R$ within the next-to-leading approximation.

Using a series representation of the amplitude which includes the dependence
on the energy scale and on the renormalization scale at subleading level,
we performed a numerical analysis in the kinematics when the two colliding photons
have the same virtuality, i.e. in the  ``pure''  BFKL regime.
We have found that the
next-to-leading order corrections coming from the kernel and from the
virtual photon to light vector meson impact factors are both large and
of opposite sign with respect to the leading order contribution.

An optimization procedure, based on the principle of minimal sensitivity method,
has proved to work nicely and has lead to stable results in the
considered energy interval, which allows us to predict the energy behavior of 
the forward amplitude. The procedure consists in evaluating
the amplitude at values of the energy parameters for which it is the least
sensitive to variations of them. We have found that there are wide regions
of values of $s_0$ and $\mu_R$ where the amplitude remains almost flat.

The optimal choices of $s_0$ and $\mu_R$ are much larger that the kinematical
scales of the problem. More than being the indication of appearance of another
scale in the problem, this could be related to the nature of the BFKL series.
The renorm- and energy scale invariance, together with the large next-to-leading
approximation corrections, call for large next-to-next-to-leading order
corrections, which are most probably mimicked by unnatural optimal values for
$s_0$ and $\mu_R$.

The use of other optimization methods and/or different, equivalent in the NLA,
representations of the amplitude gives results does not change the behaviour of
the amplitude with energy and allows for an estimate of the systematic uncertainty
of our determinations.

\end{document}